\documentstyle[12pt]{article}

\begin{document}

\begin{titlepage}
\begin{center}

\bigskip

\bigskip

\vspace{3\baselineskip}

{\Large \bf  Exact Scaling Solutions in Normal and Brans-Dicke Models of Dark
Energy}

\bigskip

\bigskip

\bigskip

\bigskip

{\bf Olga Arias\thanks{olgam@mfc.uclv.edu.cu}, Tame
Gonzalez\thanks{tame@mfc.uclv.edu.cu}, Yoelsy
Leyva\thanks{yoelsy@mfc.uclv.edu.cu}
and Israel Quiros\thanks{israel@mfc.uclv.edu.cu}}\\
\smallskip

{\small\it
Physics Department. Las Villas Central University.\\
Santa Clara 54830. Villa Clara. Cuba}

\bigskip

\bigskip

\bigskip

\bigskip

\vspace*{.5cm}

{\bf Abstract}\\
\end{center}
\noindent

A linear relationship between the Hubble expansion parameter and
the time derivative of the scalar field is explored in order to
derive exact cosmological, attractor-like solutions, both in
Einstein's theory and in Brans-Dicke gravity with two fluids: a
background fluid of ordinary matter, together with a
self-interacting scalar field accounting for the dark energy in
the universe. A priori assumptions about the functional form of
the self-interaction potential or about the scale factor behavior
are not necessary. These are obtained as outputs of the assumed
relationship between the Hubble parameter and the time derivative
of the scalar field. A parametric class of scaling quintessence
models given by a self-interaction potential of a peculiar form: a
combination of exponentials with dependence on the barotropic
index of the background fluid, arises. Both normal quintessence
described by a self-interacting scalar field minimally coupled to
gravity and Brans-Dicke quintessence given by a non-minimally
coupled scalar field are then analyzed and the relevance of these
models for the description of the cosmic evolution are discussed
in some detail. The stability of these solutions is also briefly
commented.


\bigskip

\bigskip

\end{titlepage}

\section{Introduction}

Dark energy or missing energy is one of the contemporary issues
the physics community is more interested in due, mainly, to a
relatively recent (revolutionary) discovery that our present
universe is in a stage of accelerated expansion\cite{pr}. This
missing component of the material content of the universe is the
responsible for the current stage of accelerated expansion and
accounts for 2/3 of the total energy content of the universe,
determining its destiny\cite{turner}. This is a new form of energy
with repulsive gravity and possible implications for quantum
theory and supersymmetry breaking. Among others, the new cosmology
is characterized by the following features\cite{turner}: 1) Flat,
critical density accelerating universe, 2) early period of
inflation, 3) density inhomogeneities produced from quantum
fluctuations during inflation, 4) composition of 2/3 dark energy,
1/3 dark matter and 1/200 bright stars, 5) matter content: $(29\pm
4)\%$ cold dark matter, $(4\pm 1)\%$ baryons and $\sim 0.3\%$
neutrinos. Besides, there is evidence (recent observation of
SN1997ff at redshift $z=1.7$) that the present stage of
accelerated expansion was preceded by an early period of
decelerated expansion\cite{turner}. Any theoretical framework
intended to describe our present universe should be compatible
with most of these requirements.\footnotemark\footnotetext{Most
part of these features can be considered as observational
evidence. Early inflation, on the contrary, is just a theoretical
model that is consistent with observations.}

A self-interacting, slowly varying scalar field, most often called
quintessence, has been meant to account for the dark energy
component. In the simplest models of quintessence, minimally
coupled scalar fields are considered. However, extended
quintessence models have also been
considered\cite{faraoni,uzan,dftorres}. In these models the scalar
field that accounts for the dark energy component is non-minimally
coupled to gravity and, besides, is a part of the gravity sector.
Brans-Dicke theory is a prototype for this kind. Motivated by the
possibility to have superaccelerated expansion and,
correspondingly, superquintessence solutions, this type of
theories is now being considered (see, for instance,
Ref.\cite{dftorres} and references therein).

A variety of self-interaction potentials for the quintessence
field has been studied. Among them, the simplest exponential
potential (a single exponential) model is unacceptable because it
can not produce the transition from subdominant to dominant energy
density\cite{peebles}. Other combinations of exponential
potentials have been also
studied\cite{ratrap,chimento,starobinsky,rs,tmatos}. Combinations
of exponentials are interesting alternatives since these arise in
more fundamental (particle) contexts: supergravity and
superstring\cite{bcn}, where these types of potentials appear
after dimensional reduction.

In most cases the occurrence of a self-interaction potential for
the scalar field makes difficult to solve analytically the field
equations, although some techniques for deriving solutions have
been developed. In Ref.\cite{ellis}, for instance, the form of the
scale factor is given a priori and, consequently, the
self-interaction potential can be found that obeys the field
equations. This method has been repeatedly used\cite{uggla,sen} as
well as interesting applications of it\cite{saini}. However there
are cases when exact solutions can be found once the form of the
potential is given\cite{rs,tmatos}. In other cases some suitable
relationship between the self-interaction potential and the scalar
field kinetic energy is assumed\cite{ygong}. The assumption that
the scalar field energy density scales as an exact power of the
scale factor has also been used for deriving solutions
\cite{jdbarrow}.

In case we deal with a non-minimally coupled scalar field that
accounts for the dark energy, the situation with finding exact
solutions is even more complicated. It seems that numerical
integration is the "simplest" way to treat this problem. In Ref.
\cite{dftorres}, for instance, numerical solutions are explored in
extended quintessence models.

In this paper we explore a linear relationship between the Hubble
expansion parameter and the time derivative of the scalar field to
derive exact cosmological scaling solutions\footnotemark
\footnotetext{By scaling we mean cosmological models where the
dimensionless density parameter for each component in the energy
content is a constant: $\Omega_i\equiv \rho_i/\rho_{crit}= constant$
\cite{holden}.} both in Einstein's gravity and in Brans-Dicke theory
with two fluids: a barotropic perfect fluid of ordinary matter,
together with a self-interacting scalar field fluid accounting for
the dark energy in the universe.

A similar (but more general) relationship $\dot\phi\propto H+\alpha
 L(\phi)$ has been used in Ref. \cite{chimento1} to find general
solutions in $N+1$-dimensional theories of gravity with a self-
interacting scalar field and also to derive general solutions when
one deals with two scalars: a self-interacting scalar field with
exponential potential plus a free scalar field.\footnotemark
\footnotetext{A method where a linear relationship between the
field variables and/or their derivatives is assumed has been also
used in \cite{acq} to derive 4d Poincare invariant solutions in
thick brane contexts.}

We show that the assumed linear relationship between the Hubble
parameter and the time derivative of the scalar field is suggested
by an implicit symmetry of the field equations and it can lead
always to attractor scaling solutions. In this regard, we point
out that it is not necessary to make any a priori assumptions
about the functional form of the self-interaction potential or
about the scale factor behavior. These are obtained as outputs of
the assumed relationship once one integrates the field equations
explicitly. As a result a peculiar class of potentials arise: a
combination of exponentials with some dependence on the barotropic
index of the ordinary matter fluid. This kind of potential is of
interest for models of dark energy since, as it has been pointed
out in Ref. \cite{bcn}, a single exponential potential can lead to
one of the following scaling attractor solutions: 1) a late time
attractor where the scalar field mimics the evolution of the
background barotropic fluid, and 2) the late time attractor is the
scalar field dominated solution. In consequence, a combination of
exponentials should allow for a scenario in which the universe
evolves through a radiation-matter regime (attractor 1) and at
some recent epoch evolves into the scalar field dominated regime
(attractor 2).

In this paper we will be concerned with flat
Friedmann-Robertson-Walker (FRW) cosmologies with the line element
given by:

\begin{equation}
\label{line element} ds^2= -dt^2+a(t)^2 \delta_{ik} dx^i dx^k ,
\end{equation}
where the indexes $i,k = 1,2,3$ and $a(t)$ is the scale factor. We
use the system of units in which $8\pi G=c=1$.

The paper has been organized in the following way. In section 2 we
explain the details of the method by applying it to derive
solutions in Einstein's theory with two fluids: a background fluid
of ordinary matter and a fluid of a self-interacting scalar field
that accounts for the dark energy component in the universe. We
study also, the resulting parametric class of scaling solutions
(including the space of parameters) and we briefly comment on
their stability. In particular, we make clear the point that the
assumed linear relationship between the Hubble expansion parameter
and the time derivative of the scalar field, acts just like a
selection principle that preserves only the (relevant)
attractor-like solutions. Section 3 is devoted to the derivation
of a parametric class of scaling solutions in Brans-Dicke gravity
with a background fluid and a self-interacting Brans-Dicke (BD)
field, by assuming the same linear relationship between the Hubble
expansion parameter and the time derivative of the BD field. We
apply the conformal technique to simplify the mathematical
handling of the problem. The stability of the solutions found is
also commented in this section, while their relevance for the
description of the cosmic evolution will be discussed in some
detail in section 4. Some conclusions are given in the final
section 5.

\section{The method}

In this section we explain the details of the method we use for
deriving solutions in gravity coupled to self-interacting scalars.
We will derive solutions in Einstein's gravity with two fluids: a
background barotropic fluid of ordinary matter and a
self-interacting scalar field fluid. The field equations are the
following:

\begin{equation}
\label{fec1} 3H^2=\rho_m+\frac{1}{2}\dot{\phi}^2+V,
\end{equation}

\begin{equation}
\label{fec2} 2\dot
H+3H^2=(1-\gamma)\rho_m-\frac{1}{2}\dot\phi^2+V,
\end{equation}

\begin{equation}
\label{fec3} \ddot{\phi}+3H\dot{\phi}=-V',
\end{equation}
where $V=V(\phi)$ is the self-interaction potential, $\gamma$ is
the barotropic index of the background fluid and $H=\frac{\dot
a}{a}$ is the Hubble expansion parameter. The dot accounts for
derivative in respect to the comoving time $t$ meanwhile the comma
denotes derivative in respect to the scalar field $\phi$. The
energy density of the ordinary matter (cold dark matter plus
baryons and/or radiation) is related with the scale factor through
$\rho_m=\rho_{0,\gamma} a^{-3\gamma}$, where $\rho_{0,\gamma}$ is
an integration constant coming from integrating the conservation
equation. Let us combine equations (\ref{fec1}) and (\ref{fec2})
to obtain

\begin{equation}
\label{fec4} \dot H+3H^2= \frac{2-\gamma}{2}\rho_m+V.
\end{equation}

An implicit symmetry of the left hand side (LHS) of equations
(\ref{fec3}) and (\ref{fec4})is evident under the change
$H\rightarrow k\dot\phi$. I. e., if one assumes a linear
relationship between the Hubble expansion parameter and the time
derivative of the scalar field;

\begin{equation}
\label{relation} H=k\dot\phi,\Rightarrow a=e^{k\phi},
\end{equation}
where $k$ is a constant parameter, the LHS of equations
(\ref{fec3}) and (\ref{fec4}) coincide up to the factor $k$. We
obtain a differential equation for determining the functional form
of the potential $V$:

\begin{equation}
\label{potec}
V'+\frac{1}{k}V=\frac{\gamma-2}{2k}\rho_{0,\gamma}e^{-3k\gamma\phi}.
\end{equation}

Explicit integration yields the following potential which is a
combination of exponentials:

\begin{equation}
\label{potential} V=\xi_0
e^{-\phi/k}+\frac{2-\gamma}{6k^2\gamma-2}
\rho_{0,\gamma}e^{-3k\gamma\phi},
\end{equation}
where $\xi_0$ is an integration constant. This potential is a
particular case of the potential
$V(\phi)=M^4(e^{\alpha\phi}+e^{\beta\phi})$ in Ref.\cite{bcn},
when $\alpha=1/k$, $\beta=3k\gamma$, and
$\xi_0=\frac{2-\gamma}{6k^2\gamma-2} \rho_{0,\gamma}=M^4$. It can
be given in terms of the scale factor if one considers equation
(\ref{relation}): $\label{potential1} V=\xi_0
a^{-1/k^2}+\frac{2-\gamma}{6k^2\gamma-2}\rho_{0,\gamma}a^{-3\gamma}$.

An interesting feature of this potential is that it depends on the
type of ordinary fluid which fills the universe. Otherwise, it
depends on the barotropic index $\gamma$ of the matter fluid. If
one introduces the following time variable $dt=a^{1/2k^2}d\tau$
and by substituting (\ref{potential}) back into Eq. (\ref{fec1}),
one can rewrite this last equation in the following form:

\begin{equation}
\label{hubble} (\dot
a/a)^2=\frac{2k^2\gamma}{6k^2-2}\rho_{0,\gamma}a^{-3\gamma+
1/k^2}+\frac{2k^2\xi_0}{6k^2-1},
\end{equation}
where now the dot means derivative in respect to the new time
variable $\tau$. This equation can be integrated in quadratures:

\begin{equation}
\label{integral}
\int\frac{a^{3\gamma/2-1/2k^2-1}da}{\sqrt{A+Ba^{3\gamma+1/k^2}}}=\tau+\tau_0,
\end{equation}
where $A=\frac{2k^2\gamma}{6k^2\gamma-2}\rho_{0,\gamma}$,
$B=\frac{2k^2\xi_0}{6k^2-1}$, and $\tau_0$ is an integration
constant.

\subsection{The class of solutions}

The integral in the LHS of Eq. (\ref{integral}) can be explicitly
taken to yield the following expression for the scale factor as
function of the time variable $\tau$;

\begin{equation}
\label{scalef}
a(\tau)=a_0\{\sinh[\mu(\tau+\tau_0)]\}^{2k^2/(3k^2\gamma-1)},
\end{equation}
where
$a_0=(A/B)^{k^2/(3k^2\gamma-1)}$ and
$\mu=(3k^2\gamma-1)\sqrt{B}/k^2$.
Then, according to Eq. (\ref{relation}) the scalar field can be
easily computed to yield,

\begin{equation}
\label{sf} \phi(\tau)=\frac{1}{k}\ln
a_0+\frac{2k}{3k^2\gamma-1}\ln\sinh[\mu(\tau+\tau_0)].
\end{equation}

The Hubble expansion parameter can be computed by the knowledge of
the scale factor (\ref{scalef}):

\begin{equation}
\label{hubble1} H(\tau)=\sqrt{\frac{2k^2}{6k^2-1}\xi_0}
a(\tau)^{-1/2k^2}\coth[\mu(\tau+\tau_0)].
\end{equation}

For the purpose of observational testing of the solutions it is
useful to look for another magnitudes of astrophysical interest.
Among them, the matter density parameter:

\begin{equation}
\label{omegam}
\Omega_m=\rho_m/3H^2=\frac{3k^2\gamma-1}{3k^2\gamma}\{\cosh[\mu
(\tau+\tau_0)]\}^{-2},
\end{equation}
the scalar field energy density and scalar field density
parameter:

\begin{eqnarray}
\label{omegaphi}
\rho_\phi=\frac{1}{2}\dot\phi^2+V&=&3H^2-\rho_m,\nonumber\\
\Omega_\phi=\frac{\rho_\phi}{3H^2}&=&1-\Omega_m,
\end{eqnarray}
respectively. The second equation in (\ref{omegaphi}) is just
another writing for the field equation (\ref{fec1}). The scalar
field state parameter can be written in the following form:

\begin{equation}
\label{statep}
\omega_\phi=\frac{\frac{1}{2}\dot\phi^2-V}{\frac{1}{2}\dot\phi^2+V}=-1+
\frac{1}{3k^2(1-\Omega_m)},
\end{equation}
meanwhile the deceleration parameter $q=-(1+\dot H/H^2)$:

\begin{equation}
\label{deceleration}
q=-1+\frac{1}{2k^2}+\frac{3\gamma}{2}\Omega_m.
\end{equation}

While deriving equations (\ref{omegaphi})-(\ref{deceleration}) we
have used the field equations (\ref{fec1}), (\ref{fec2}),
(\ref{fec3}) and their combinations. Assuming the linear
relationship (\ref{relation}) between the Hubble parameter and the
time derivative of the scalar field means that we are introducing
a new free parameter, however, as it will be shown latter on in
subsection 2.B, the space of parameters can be reduced by the
combined action of choosing an appropriated normalization and
considering some well established observational facts. For purpose
of observational testing of the solution it is desirable to have
the physical parameters of observational interest written as
functions of redshift $z$. This is achieved simply by replacing in
the former equations

\begin{equation}
\label{omegamz}
\Omega_m(\tau)\rightarrow\Omega_m(z)=(\frac{\rho_{0,\gamma}}{3A})
\frac{(1+z)^{2n}}{(1+z)^{2n}+B/A},
\end{equation}
where $2n=3\gamma-1/k^2$. While deriving this expression one
has considered $a(z)=a(0)/(z+1)$, and the normalization $a(0)=1$.
In terms of the redshift the Hubble expansion parameter can be written
as:

\begin{equation}
\label{hubblez}
H(z)=\sqrt{A}\sqrt{(z+1)^{3\gamma}+\frac{B}{A}(z+1)^{1/k^2}},
\end{equation}
meanwhile, the background energy density $\rho_m=\rho_{0,\gamma}(z+1)^{3\gamma}$.

\subsection{The space of parameters}

We now investigate the space of parameters
($\gamma,\xi_0,\rho_{0\gamma},k$) parameterizing the solutions (the
parameter $\gamma$ can be fixed to be 1 at present and $4/3$ in
the past). We consider $k^2>0$ since otherwise the scalar field
would be a complex one. We should differentiate two ranges. For
$0<k^2<1/3\gamma$ the scale factor evolves periodically:

\begin{equation}
\label{scalef2} a(\tau)=\tilde a_0
\{\sin[\tilde\omega(\tau+\tau_0)]\}^{2k^2/(3k^2\gamma-1)},
\end{equation}
where $\tilde
a_0=(-A/B)^{k^2/(3k^2\gamma-1)}$. However, in this case, for
$0\leq\gamma\leq 2$, $3k^2\gamma-1<0$ so one should have
$\xi_0<0$. This means that the self-interaction potential itself
is negative, i. e., the scalar field energy density is negative.
Besides $q>0$ always and this case is not of observational
interest. Therefore we drop this interval in parameter $k$.

For $k^2>1/3\gamma$ the equations (\ref{scalef})-(\ref{hubble1})
are valid in that form. To constrain further the space of
parameters we use some observational facts. However, this point is
left for further discussion in section 4. We now determine the
constant $A/B$. For this purpose one uses two observational facts:
1) Since, at present $\Omega_m(0)=1/3$, then one obtains from Eq.
(\ref{omegamz}) that:

\begin{equation}
\label{ab1} A+B=\rho_{0,\gamma},
\end{equation}
and 2) Since radiation-matter dominated the early stages of the
evolution, then one may require that
$\Omega_m(\infty)=1-\varepsilon$, where the new parameter
$\varepsilon << 1$. This small parameter is necessary since a
matter-dominated ($\Omega_m=1$) is not stable \cite{clw}. In consequence,
from the same Eq.(\ref{omegamz}), one obtains that:

\begin{equation}
\label{ab2}\rho_{0,\gamma}=3A(1-\varepsilon).
\end{equation}

Combining the above equalities one gets that:
$\frac{B}{A}=2-3\varepsilon$. In consequence:

\begin{equation}
\Omega_m(z)=(1-\varepsilon)\frac{(z+1)^{2n}}{(z+1)^{2n}+2-3\varepsilon},
\end{equation}
and,
\begin{equation}
H(z)=\sqrt{\rho_{o,\gamma}}\sqrt{\frac{(z+1)^{3\gamma}+(2-3\varepsilon)(z+1)^
\frac{1}{k^2}}{3(1-\varepsilon)}}.
\end{equation}

One can reduce the space of parameters
$(\gamma,\varepsilon,\rho_{0,\gamma},k)$ by requiring the
following normalization: $H(0)=1$, implying that
$\rho_m(0)=\rho_{0,\gamma}=1$. We are then left with a
three-parameter class of solutions spanned by
$(\gamma,\varepsilon,k)$. In section 4 we will further constrain
the parameters $\varepsilon$ and $k$ by the observations.

\subsection{Stability of the solutions}

The stability analysis of the model given by Eqs.(2.1)-(2.3) has
been performed in Ref.\cite{clw} for the class of exponential
potentials of the kind: $V=V_0 e^{-\lambda\phi}$  $(k=8\pi G=1)$.
There are up to five fixed points (critical points). Two of them
are late-time attractor solutions. One of these is the scalar
field dominated solution: $\Omega_\phi=1$, with an affective
barotropic index $\gamma_\phi=\lambda^2/3$. This solution exist
for $\lambda^2<6$, and is a lite-time attractor in the presence of
a background fluid when $\lambda^2<3\gamma$ \cite{clw}. The other
attractor solution occurs for $\lambda^2>3\gamma$ and neither the
scalar-field nor the background fluid entirely dominates the
evolution.It is a scaling solution where
$\Omega_\phi=\frac{3\gamma}{\lambda^2}$. At this point we want to
discuss on the implications of the linear relationship (2.5)
assumed in order to find exact solutions. The procedure for the
stability analysis \cite{clw} implies writing the evolution
equations (2.1)-(2.3) as a plane-autonomous system. For this
purpose, one should define new variables:

\begin{equation}
x\equiv\frac{\dot\phi}{\sqrt6H},\quad  y\equiv\frac{\sqrt
V}{\sqrt3H}.
\end{equation}

Considering (\ref{relation}), the first equation implies
$x=\frac{1}{k\sqrt6}$. Meanwhile the critical points leading to
the former attractor solutions are $x_1=\lambda_1/\sqrt6$ for the
scalar dominated solution and
$x_2=\sqrt\frac{3}{2}\gamma/\lambda_2$ for the scaling solution
\cite{clw}. On the other hand, our potential
(Eq.(\ref{potential})) contains a combination of exponentials with
exponents $\lambda_1=1/k$ and $\lambda_2=3k\gamma$. We see then,
that assuming the linear relationship (\ref{relation}) means
dropping of the critical points but the ones that can lead to
attractor solutions. Otherwise, relationship (\ref{relation})
could be written in the following equivalent way:
$\gamma_\phi\Omega_\phi=\frac{1}{3k^2}$. Then one is faced with
two possibilities: either
$\Omega_\gamma=1\Rightarrow\gamma_\phi=\frac{1}{3k^2}$ (scalar
field dominated solution), or $\gamma_\phi=\gamma,\quad
\Omega_\phi=\frac{1}{3k^2\gamma}=const$ (scaling solution). In
this sense one can say that the assumed relationship between $H$
and $\dot\phi$ (Eq. (\ref{relation})) acts just like a selection
principle that, among the possible critical points, preserves only
the (relevant) attractor-like solutions.

\section{Brans-Dicke quintessence}

Now we use the former method to derive exact solutions in
Brans-Dicke gravity with a self-interacting scalar field. As
mentioned in the introduction to this paper, this is one in a kind
of models of extended quintessence studied before in
\cite{faraoni,uzan,dftorres}. In Ref. \cite{dftorres}, in
particular, numerical solutions in extended models of quintessence
(including Brans-Dicke) were explored.

The Brans-Dicke field equations, in the presence of a
self-interacting scalar field, are:

\begin{equation}
\label{bdfeq1}
3H^2+3H\dot\varphi-\frac{\omega}{2}\dot\varphi^2=e^{-\varphi}(\rho_m+V),
\end{equation}

\begin{equation}
\label{bdfeq2} 2\dot
H+3H^2+\ddot\varphi+\dot\varphi^2+2H\dot\varphi+\frac{\omega}{2}
\dot\varphi^2=e^{-\varphi}\{(1-\gamma)\}\rho_m+V,
\end{equation}

\begin{equation}
\label{bdfeq3}
\ddot\varphi+\dot\varphi^2+3H\dot\varphi=\frac{e^{-\varphi}}{2\omega+3}
\{(4-3\gamma)\rho_m+4V-2V'\},
\end{equation}
where we have introduced a new field variable $\varphi$ instead of
the original $\phi=e^\varphi$. In these equations $\omega$ is the
Brans-Dicke coupling parameter, $\gamma$ is the barotropic index
of the fluid of ordinary matter, $V$ is the self-interaction
potential and $H=\frac{\dot a}{a}$ is the Hubble expansion
parameter. The dot accounts for derivative in respect to the
comoving time $t$ meanwhile the comma denotes derivative in
respect to the scalar field $\varphi$. As before, the energy
density of the ordinary matter (cold dark matter plus baryons
and/or radiation) is related with the scale factor through
$\rho_m=\rho_{0,\gamma} a^{-3\gamma}$, where $\rho_{0,\gamma}$ is
an integration constant coming from integrating the conservation
equation.

For simplicity of the mathematical handling of the problem it is
convenient to perform a conformal transformation of the metric:
$\bar g_{ab}= e^{\psi/\sqrt{\xi}} g_{ab}$, where $\xi=\omega+3/2$
and, for convenience, we have rescaled the BD field
$\varphi=\psi/\sqrt{\xi}$. The transformed line element is now:
$d\bar s^2=-d\bar t^2+\bar a^2 \delta_{ij}dx^i dx^j$, where the
conformal comoving time and scale factor are related with the
original (Jordan frame) ones through:
$dt=e^{-\psi/(2\sqrt{\xi})}d\bar t$, and
$a=e^{-\psi/(2\sqrt{\xi})}\bar a$ respectively. Then we will be
faced with the Einstein frame (EF) formulation of BD theory. The
relevant field equations are now the following:

\begin{equation}
\label{effeq1} 3\bar H^2=\bar\rho_m+\bar\rho_\psi,
\end{equation}

\begin{equation}
\label{effeq2} 2\dot{\bar H}+3\bar H^2=(1-\gamma)\bar\rho_m-\bar
P_\psi,
\end{equation}

\begin{equation}
\label{effeq3} \ddot\psi+3\bar H
\dot\psi=\frac{4-3\gamma}{2\sqrt{\xi}}\bar\rho_m - \bar V',
\end{equation}
where the EF scalar field energy density
$\bar\rho_\psi=\frac{1}{2}\dot\psi^2+\bar V$ and the corresponding
pressure $\bar P_\psi=\frac{1}{2}\dot\psi^2-\bar V$. Besides, the
transformed "conservation" equation is:

\begin{equation}
\label{efcons} \dot{\bar\rho}_m+(3\gamma\bar H+\sqrt\frac{2}{3}
W\dot\psi)\bar\rho_m=0,
\end{equation}
where the "exchange" function
$W=\sqrt\frac{3}{2}\frac{4-3\gamma}{2\sqrt{\xi}}$ has been
introduced as in Ref.\cite{holden}. After integration:
$\bar\rho_m=\bar\rho_{0,\gamma}\bar
a^{-3\gamma}e^{-\sqrt\frac{2}{3} W\psi}$. Summing up
(\ref{effeq1}) and (\ref{effeq2}) yields:

\begin{equation}
\label{constrain} \dot{\bar H}+3\bar
H^2=\frac{2-\gamma}{2}\bar\rho_m+\bar V.
\end{equation}

As before an implicit symmetry of the LHS of equations
(\ref{effeq3}) and (\ref{constrain}) is made explicit by the
change $\dot\psi\rightarrow\lambda\bar H$. In this case
Eq.(\ref{effeq3})$\rightarrow$ $\lambda$ times
Eq.(\ref{constrain}). Therefore, if we assume the linear
relationship

\begin{equation}
\label{efrel} \dot\psi=\lambda\bar
H,\;\;\Rightarrow\;\;e^\psi=\bar a^\lambda,
\end{equation}
then the following differential equation for the self-interaction
potential $V$ is obtained: $\bar V'+\lambda\bar
V=[2\sqrt\frac{2}{3}W-(2-\gamma)\lambda]\bar\rho_m/2$.
Straightforward integration of this equation yields:

\begin{equation}
\label{efpotential} \bar V(\psi)=\bar V_0 e^{-\lambda\psi}+ \bar
W_0 e^{-\delta\psi},
\end{equation}
where $\delta=3\gamma/\lambda+\sqrt{2/3}W$, $\bar V_0$ is an
integration constant and $\bar
W_0=[\sqrt\frac{2}{3}W-\frac{2-\gamma}{2}\lambda]\bar\rho_{0,\gamma}/(\lambda-\delta)$.
In terms of the scale factor Eq. (\ref{efpotential}) reads: $\bar
V(\psi(\bar a))=\bar V_0 \bar a^{-\lambda^2}+\bar W_0 \bar
a^{-\delta\lambda}$.

This potential belongs to the same class that (\ref{potential})
and shows the interesting feature that it depends on the type of
ordinary fluid which fills the universe (it depends on the
barotropic index $\gamma$ of the matter fluid). By substituting
Eq. (\ref{efpotential}) back into the Friedmann equation
(\ref{effeq1}) one obtains

\begin{equation}
\label{friedmann} \bar H^2=A \bar a^{-\lambda\delta}+B \bar
a^{-\lambda^2},
\end{equation}
where
$A=(\lambda^2-6)\gamma/(\lambda^2-\sqrt{2/3}W\lambda-3\gamma)$,
$B=2\bar V_0/(6-\lambda^2)$, and we have considered, besides, that
$\bar\rho_m=\bar\rho_{0,\gamma} \bar
a^{-3\gamma-\sqrt\frac{2}{3}W\lambda}=\bar\rho_{0,\gamma} \bar
a^{-\delta\lambda}$.

To solve (\ref{friedmann}) we make another change of time variable:
$dr=\bar a^{-\lambda^2/2} d\bar t$ to get Eq. (\ref{friedmann})
integrated in quadratures:

\begin{equation}
\label{quadrature} \int\frac{\bar
a^{\lambda(\delta-\lambda)/2-1}d\bar a}{\sqrt{\frac{A}{B}+ \bar
a^{\lambda(\delta-\lambda)}}}=\sqrt{B}(r+r_0),
\end{equation}
where $r_0$ is another integration constant.

\subsection{The class of solutions}

Straightforward integration of Eq. (\ref{quadrature}) leads to:

\begin{equation}
\label{efscalef} \bar
a(r)^{\lambda(\delta-\lambda)/2}=\sqrt{A/B}\sinh[\mu(r+r_0)],
\end{equation}
with $\mu=\sqrt{B/\xi}\lambda(\delta-\lambda)/2$. For the rescaled
BD scalar field we obtain:

\begin{equation}
\label{efsf}
e^\psi=\{\sqrt{A/B}\sinh[\mu(r+r_0)]\}^{2/(\delta-\lambda)}.
\end{equation}

In consequence the dimensionless background density parameter:

\begin{equation}
\label{efbdp} \bar\Omega_m=\bar\rho_m/3\bar
H^2=\frac{\bar\rho_{0,\gamma}}{3A}\frac{\bar
a^{\lambda(\lambda-\delta)}}{\bar
a^{\lambda(\lambda-\delta)}+A/B},
\end{equation}

and,

\begin{equation}
\label{efsfdp} \bar\Omega_\psi=1-\bar\Omega_m.
\end{equation}

As before it is useful to have the magnitudes of observational
interest written in terms of the redshift $z$. In the Jordan frame
we have that $a(z)=a(0)/(z+1)$. We choose the normalization
$a(0)=1\;\rightarrow\; \bar a(0)=1$. Then, since
$a=e^{-\psi/(2\sqrt\xi)}\bar a=\bar a^n\; \rightarrow\; \bar
a=(z+1)^{-1/n}$, where we have introduced the following definition
for the constant $n$:

\begin{equation}
\label{n} n=\frac{2\sqrt\xi-\lambda}{2\sqrt\xi}.
\end{equation}

After these considerations the dimensionless density parameter and the
Hubble expansion parameter can be written in the following way:

\begin{equation}
\label{efbdpz}
\bar\Omega_m(z)=\frac{\bar\rho_{0,\gamma}}{3A}\frac{(z+1)^
{(\delta\lambda-\lambda^2)/n}}{(z+1)^{(\delta\lambda-\lambda^2)/n}+B/A},
\end{equation}
and,

\begin{equation}
\label{efhubblez} \bar H(z)=\sqrt A \sqrt{(z+1)^{\lambda\delta/n}+\frac{B}
{A}(z+1)^{\lambda^2/n}},
\end{equation}
respectively. Other physical magnitudes of observational interest are the
scalar field equation of state (and barotropic index):

\begin{equation}
\label{efstateq} \bar\omega_\psi=-1+\bar\gamma_\psi,\;\;\bar\gamma_\psi=
\frac{\lambda^2}{3(1-\bar\Omega_m)},
\end{equation}
and the deceleration parameter: $\bar q=-(1+\dot{\bar H}/\bar H^2)$;

\begin{equation}
\label{efdecel} \bar q=-1+\lambda^2/2+(3\gamma/2)\bar\Omega_m.
\end{equation}

\subsubsection{Relationship among frames}

Now it is useful to give the main physical magnitudes of
observational interest also in the Jordan frame, as related with
the EF magnitudes. Recall that $dt=e^{-\psi/2\sqrt\xi}d\bar t=
\bar a^{-\lambda/(2\sqrt\xi)}d\bar t$, $a=e^{-\psi/2\sqrt\xi}\bar
a=\bar a^n$ (see the definition for $n$ in Eq. (\ref{n})), and
$\bar
T_{ab}=e^{-\psi/\sqrt\xi}T_{ab}\;\rightarrow\;\rho_m=e^{2\psi/\sqrt\xi}
\bar\rho_m=\bar a^{2\lambda/\sqrt\xi}\bar\rho_m$. Other important
relationships are the Hubble parameter:

\begin{equation}
\label{efjfhubble} H=ne^{\psi/(2\sqrt\xi)}\bar H=n\bar a^{\lambda/
(2\sqrt\xi)}\bar H,
\end{equation}
the dimensionless density parameter,

\begin{equation}
\label{efjfdp} \Omega_m=(1/n^2)\bar\Omega_m,
\end{equation}
the dimensionless scalar field density parameter,\footnotemark
\footnotetext{In the Jordan frame we define
$\Omega_i=\rho_i/\rho_{critical}$, where $\rho_{critical}=3e^\phi
H^2$, so that the Friedmann constrain equation (\ref{bdfeq1}) can
be written as: $1=\Omega_m+\Omega_\phi$.}

\begin{equation}
\label{efjfsfdp} \Omega_\phi=1-(1/n^2)\bar\Omega_m,
\end{equation}
the scalar field equation of state (and the barotropic index):

\begin{equation}
\label{efjfstateq}
\omega_\phi=-1+\gamma_\phi,\;\;\gamma_\phi=\lambda\frac{(n/3)(\lambda-1/
\sqrt\xi)-\frac{1}{2}(\gamma/\sqrt\xi)\bar\Omega_m}{n^2-\bar\Omega_m},
\end{equation}
and the deceleration parameter,

\begin{equation}
\label{efjfdecel}
q=(1/n)[\bar q+(1-n)(1-\lambda^2/2)-(n/2)\lambda/\sqrt\xi].
\end{equation}

This class of solutions depends (in principle) on 5 parameters: $\lambda$,
$\xi$, $\bar\rho_0$, $\bar V_0$, and $\gamma$. The constants $A$ and $B$
are given in terms of the aforementioned parameters.

\subsection{The space of parameters}

In this subsection we will reduce the space of parameters by
properly choosing the normalization for the Hubble parameter and
by using some observational facts. In the Einstein frame the
Friedmann constrain equation (\ref{effeq1}) means that
$\bar\Omega_m\leq 1$. Besides, the solution $\bar\Omega_m=1$ is
not stable so we are left with $\bar\Omega_m<1$. The same is true
in the Jordan frame, where the Friedmann constrain equation
(\ref{bdfeq1}) means that $0\leq\Omega_m\leq 1$. On the one hand,
$\Omega_m$ is a maximum at $z=\infty$ (the same is true for
$\bar\Omega_m$). Then, we require that
$\Omega_m(\infty)=1-\varepsilon$, where, as before, $\varepsilon$
is a small number. This, in turn, yields:
$\bar\Omega_m(\infty)=n^2(1-\varepsilon)$ In consequence,
$\bar\rho_{0,\gamma}=3n^2A(1-\varepsilon)$. On the other hand,
according to the observations, at present (in the JF)
$\Omega_m(0)=(1/n^2)\bar\Omega_m(0)=1/3$. This last equality
implies that $\bar\rho_{0,\gamma}=n^2(A+B)$. Combining of the
above constrains yields $B/A=2-3\varepsilon$. In consequence, Eq.
(\ref{efbdpz}) and Eq. (\ref{efhubblez}) can be written in the
following way:

\begin{equation}
\label{efbdpz1} \bar\Omega_m(z)=n^2(1-\varepsilon)\frac{(z+1)^
{(\delta\lambda-\lambda^2)/n}}{(z+1)^{(\delta\lambda-\lambda^2)/n}+
2-3\varepsilon},
\end{equation}
and,

\begin{equation}
\label{efhubblez1} \bar H(z)=\frac{\sqrt{\bar\rho_{0,\gamma}}}{n}
\sqrt{\frac{(z+1)^{\lambda\delta/n}+(2-3\varepsilon)
(z+1)^{\lambda^2/n}}{3(1-\varepsilon)}},
\end{equation}
respectively. The space of parameters
($\gamma,\lambda,\varepsilon,\bar\rho_{0,\gamma}$) can be further
reduced by choosing the normalization in which
$H(0)=1\;\Rightarrow\;\bar
H(0)=1/n\;\Rightarrow\;\bar\rho_{0,\gamma}=1$. In this case we are
left with a three-parametric class of solutions spanned by the
free parameters ($\gamma,\lambda,\varepsilon$). In section 4 we
will further constrain the free parameters $\lambda$ and
$\varepsilon$ by means of comparison with supernovae observational
data.

\subsection{Existence and Stability of the Solutions}

The stability of the system of dynamical equations
(\ref{effeq1}-\ref{effeq3}) has been already studied in
(\cite{holden}) for a single exponential of the kind $\bar V= \bar
V_0 e^{-g\psi}$. This study has been performed in the Einstein
frame, however, the existence and stability of critical points
already defined in the Einstein frame, remain unchanged under a
conformal transformation back to the Jordan frame \cite{holden}.
The system (\ref{effeq1}-\ref{effeq3}) can be arranged in the form
of a plane-autonomous system if one introduces the variables:

\begin{equation}
\label{var1} x=\frac{\dot\psi}{\sqrt6\bar H},\quad
y=\frac{\sqrt{\bar V}}{\sqrt3\bar H}.
\end{equation}

In this case one obtains \cite{holden}:
\begin{equation}
\label{ps1}
y'=-\sqrt\frac{3}{2}gxy+\frac{3}{2}y(2x^2+\gamma(1-x^2-y^2)),
\end{equation}
and,

\begin{equation}
\label{ps2}
x'=-3x+\sqrt\frac{3}{2}gy^2+\frac{3}{2}x(2x^2+\gamma(1-x^2-y^2))+W(1-x^2-y^2).
\end{equation}

Following our previous result in section 2.C, that the linear
relationship between the time derivative of the scalar field and
the Hubble parameter leads to solutions being attractors of two
possible kinds: scalar field dominated and scaling solutions, then
one expects that the relationship (\ref{efrel}) leads to the same
result in the present situation.

The system described by (\ref{ps1}) and (\ref{ps2}) can have up to
five fixed points depending on the values of $g,\gamma$ and $W$
\cite{holden}. However, when combined with the relationship
(\ref{efrel}), that leads to the fixed value: $x=\lambda/\sqrt6$,
it leaves us with two possibilities (see the potential in Eq.
(\ref{efpotential})): 1) Either
$g_1=\lambda\Rightarrow\bar\Omega_\psi=1$, and
$\bar\gamma_\psi=\lambda^2/3$, or 2)
$g_2=\delta=3\gamma/\lambda+\sqrt{2/3}W\Rightarrow\bar\Omega_\psi=
\lambda(\lambda-\sqrt{2/3}W)/3\gamma$, and
$\bar\gamma_\psi=\gamma\lambda/(\lambda-\sqrt{2/3}W)$.

In the first case, in the Einstein frame, we have a scalar field
dominated solution, meanwhile, in the second case we have a
scaling solution. Solution 1) exists for $\lambda^2<6$ and is
stable when
$W-\sqrt{W^2+18\gamma}<\sqrt6\lambda<W+\sqrt{W^2+18\gamma}$.
Meanwhile, solution 2) exists for $\lambda>3\gamma
W/(\frac{9\gamma(2-\gamma)}{2\sqrt6}-1)$ and either
$\lambda>-\frac{3\sqrt6\gamma}{W+\sqrt{W^2+18\gamma}}$ or
$\lambda<-\frac{3\sqrt6\gamma}{W-\sqrt{W^2+18\gamma}}$. It is
stable whenever it exists \cite{holden}.

\section{Observational Testing}

Although we can not talk literally about observational testing of
the solutions found (we point out that the observational facts are
not yet conclusive in some cases and are not accurate enough in
others), in this section we shall study whether our solutions
agree with some observational constrains that are more or less
well established. The main observational facts we consider are the
following\cite{turner}:

1.- At present ($z=0$) the expansion is accelerated ($q(0) < 0$).

2.- The accelerated expansion is a relatively recent phenomenon.
Observations point to a decelerated phase of the cosmic evolution
at redshift $z=1.7$. There is agreement in that transition from
decelerated into accelerated expansion occurred at a $z \approx
0.5$ \cite{triess}.

3.- The equation of state for the scalar field at present
$\omega_\phi(\omega_\varphi)\sim -1$ (it behaves like a
cosmological constant). With a 95$\%$ confidence limit
$\omega<-0.6$\cite{ptw}.

4.- Although, at present, both the scalar (quintessence) field and
the ordinary matter have similar contributions in the energy
content of the universe ($\Omega_m(0)=1/3\;
\Rightarrow\;\Omega_\phi(0)=2/3$), in the past, the ordinary
matter dominated the cosmic evolution,
\footnotemark\footnotetext{A sufficiently long matter dominated
decelerated phase is needed for the observed structure to develop
from the density inhomogeneities \cite{sen}} meanwhile, in the
future, the quintessence field will dominate (it already
dominates) and will, consequently, determine the destiny of the
cosmic evolution.

Now we proceed to "observationally" test the solutions found in
the cases studied in the former sections.

\subsection{Minimally coupled quintessence}

In the case of normal quintessence model studied in section 2, our
solution depends on two parameters $\varepsilon$ and $k$ (we fix
$\gamma=1$, meaning cold dark matter dominance at present). To
constrain the values of these parameters we use the following
facts:

1. Corasaniti and Copeland \cite{cc} found that the determination
of the third peak in the BOOMERANG data limits the value today of
equation of state $-0.93 \leq \omega_{\varphi}(0) \leq -1$

2. As stated before \cite{bhm}, nucleosynthesis predictions claims
that at $95\%$  confidence level. $\Omega_{\varphi} (1 MeV \simeq
z = 10^{10}) \leq 0.45$

3. During galaxy formation epoch \cite{mst}, around $z \approx
2-4$, the value of quintessence density parameter is
$\Omega_\varphi< 0.5$

Using these constrains we bound the parameter space
($k,\varepsilon$) by means of a computing code. These parameters
range as $0\leq\varepsilon\leq 0.045$ and $k > 0$.

Now we select a pair from the above interval in order to compare
our model with Supernova data. We first compare our model with the
theoretical predictions of the Lambda Cold Dark Matter model
($\Lambda CDM$) for modulus distance. Subsequently, we follow the
procedure used in Ref. \cite{rubano} to further test our model.
However, this procedure, which implies direct comparison of our
model with observations of effective apparent magnitudes of SNIa,
can not be used to further constrain the space of parameters,
since the free parameters in our model are not much sensible to
these observations. For instance if we consider
$\varepsilon=0.01$, the $\chi^2$ distribution has a minimum in
$m_0$ direction\footnotemark\footnotetext{$m_0$ is a parameter
connected to the absolute magnitude and the Hubble parameter} at
$m_0=23.97$ ($\chi^2=62.5996$ or $\chi^2/dof=1.18$ per degree of
freedom) , however it has no minimum in the $k$ direction, meaning
that the parameter $k$ can take any positive
value.\footnotemark\footnotetext{In the limit $k\to\infty$ the
$\Lambda$CDM model is recovered.} In fig. 1, $\chi^2$ is plotted
as function of the free parameters $k$ and $m_0$ (we chose
$\varepsilon=0.01$, $k$ could be any value in the physically
meaningful range). Although the model fulfils the observational
requirements $1-4$ given at the beginning of this section, we
point out that other observations should be considered in order to
further constrain the space of parameters.

\subsection{Brans-Dicke quintessence}

In this case our solution depends on two parameters $\lambda$ and
$\varepsilon$ (recall that we fix $\gamma$ to be unity). We apply
exactly the same procedure that in the former subsection to
constrain the parameter space. However, in this case we should
differentiate among the two relevant frames: The JF and the EF, in
which Brans-Dicke theory can be formulated. We treat both
formulations separately, bearing in mind that none of them could
be taken to be the one with real physical meaning. We recall that,
given the uncertain character of the dark energy (it dominates the
energy content of the universe) and dark matter (it dominates the
matter density in the universe), it could be possible that the
dark energy and the dark matter, on the one side, and the ordinary
(barionic) matter on the other, are minimally coupled with respect
to different metrics (see Ref.\cite{holden} and references
therein). Therefore, both conformal formulations are of physical
significance.

When the aforementioned procedure is applied to the Einstein frame
magnitudes of observational interest, one finds that the
physically meaningful region in parameter space is bounded by
$0\leq \varepsilon \leq 0.045$ and $0 < \lambda < 0.37$.
Meanwhile, in the Jordan frame one obtains that the physically
meaningful range is $0\leq \varepsilon \leq 0.045$ and $0.03 <
\lambda < 0.38$.

It should be pointed out that, in all cases, the physically
meaningful region in parameter space is chosen such that the main
observational facts $1-4$ explained at the beginning of the
section are fulfilled.  As an illustration, in fig. 2, we show the
evolution of both dimensionless background and scalar field energy
densities $\&$ $z$ for the Einstein frame magnitudes
$\bar\Omega_m$ and $\bar\Omega_\psi$
respectively\footnotemark\footnotetext{The Jordan frame magnitudes
$\Omega_m$ and $\Omega_\varphi$ do not differ significantly from
the corresponding EF ones.} (the values of the free parameters are
taken from the physically meaningful region in parameter space).
In fig. 3, the evolution of the equation of state in both
formulations $\bar\omega_\psi$ and $\omega_\varphi$ $\&\;z$ is
shown, meanwhile, in fig. 4, we plot the deceleration parameter in
both frames $\bar q(z)$ and $q(z)$ to show the transition redshift
when the evolution turns from decelerated into accelerated
($z*=0.54$ and $z*=0.56$ in the EF and in the JF respectively). It
is seen that the differences in these magnitudes are not
significant as one goes from one frame into the other. In fig. 5,
as an illustration, we plotted the effective apparent magnitude
observed \cite{perlmutter} and calculated according to our model
with parameters ($\varepsilon =0.01$,$\lambda=0.2$) to show the
good fitting (the results are within 1$\sigma$ expectation).

We want to stress that, as in the former case (subsection 4.1),
the free parameters of the model are not much sensible to the SNIa
data so, the standard procedure used for instance in Ref.
\cite{rubano}, is not suitable for further constraining of the
space of parameters of the model. Other observational evidence
could be considered for this purpose.

\section{Conclusions}

We have found a new parametric class of exact cosmological scaling
solutions in gravity theory minimally coupled to a
self-interacting scalar field  and in Brans-Dicke theory with a
self-interacting BD scalar, by assuming a linear relationship
between the Hubble expansion parameter and the time derivative of
the scalar field. This relationship is suggested by an implicit
symmetry of the field equations.

We point out that it is not necessary to make any a priori
assumptions about the functional form of the self-interaction
potential or about the scale factor behavior. These are obtained
as outputs of the assumed relationship once one integrates the
field equations explicitly. As a result a peculiar class of
potentials arise: a combination of exponentials with some
dependence on the barotropic index of the ordinary matter fluid.
This kind of potential is of interest for models of dark energy.
Recall that a single exponential potential can lead to one of the
following scaling attractor solutions\cite{bcn}: 1) a late time
attractor where the scalar field mimics the evolution of the
background barotropic fluid, and 2) the late time attractor is the
scalar field dominated solution. In consequence, a combination of
exponentials should allow for a scenario in which the universe
evolves through a radiation-matter regime (attractor 1) and at
some recent epoch evolves into the scalar field dominated regime
(attractor 2).

By inspection of the stability and existence of the solutions
found, we show that the assumed linear relationship between the
Hubble parameter and the time derivative of the scalar field can
lead always (both in standard minimally coupled and in Brans-Dicke
models of dark energy) to attractor scaling solutions. Otherwise,
the assumed relationship acts as a selection principle: only two
of five critical points in phase space agree with it.

In all cases, the relevance of the solutions found is outlined and
the space of parameters is constrained with the help of a computed
code that considers some well established observational facts. The
models explored here are tested with the SNIa observational data.
Agreement with observations is always within $1\sigma$
expectation. It is pointed out, however, that the procedure of
Ref. \cite{rubano}, can not be used to further constrain the
parameter space, since the free parameters in the models of
interest are not sensible to the luminosity distance SNIa
observational data. Nevertheless, the scaling solutions found in
both cases, meet the main observational facts considered in this
paper with acceptable accuracy (points 1-4 at the beginning of
section 4.). It is also pointed out that other observational data
should be considered to further constrain the space of parameters.

In the case we deal with minimally coupled quintessence, the model
explored in this paper is a particular case of the one in
Ref.\cite{bcn}. However, in that paper the authors did not
consider any exact solutions at all but their aim was to
numerically explore the model. In the Brans-Dicke case, as far as
we know, the combination of exponentials has not been explored
yet, although the stability and existence of scaling solutions in
Brans-Dicke quintessence with a single exponential potential has
been studied, for instance, in Ref.\cite{holden}.


We thank our colleagues Rolando Cardenas, Diosdado Villegas and
Carlos R. Fadragas for helpful comments and Indrajit Chakrabarty
for pointing to us Ref.\cite{holden}. We acknowledge the MES of
Cuba by financial support of this research.


\newpage

\begin{figure}[b]
\caption{The function $\chi^2$ is plotted as function of the free
parameters $k$ and $m_0$ (we chose $\varepsilon=0.01$. As seen $k$
could be any value in the physically meaningful range so, SNIa
luminosity observations do not allow for further constrain of the
parameter space. Other observations could be considered for this
purpose.}
\end{figure}

\begin{figure}[b]
\caption{We show the evolution of both dimensionless background
and scalar field energy densities $\&$ $z$ for the Einstein frame
magnitudes $\bar\Omega_m$ and $\bar\Omega_\psi$. The parameters
$\lambda=0.2$ and $\varepsilon=0.01$ have been chosen so that the
SNIa data is reproduced within $1\sigma$. Equality of matter and
quintessence energy density occurs approximately at $z\approx
0.3-0.4$.}
\end{figure}

\begin{figure}[b]
\caption{The evolution of the equation of state in both
formulations $\bar\omega_\psi$ and $\omega_\varphi$ $\&\;z$ is
shown. The equation of state for the quintessence field in the
Einstein frame is never positive.}
\end{figure}

\begin{figure}[b]
\caption{We plot the deceleration parameter in both frames $\bar
q(z)$ and $q(z)$ to show the transition redshift when the
evolution turns from decelerated into accelerated: $z*=0.54-0.56$.
It is seen that the differences in these magnitudes are not
significant as one goes from one frame into the other.}
\end{figure}

\begin{figure}[b]
\caption{As an illustration, we plotted the effective apparent
magnitude observed \cite{perlmutter} and calculated according to
our model with parameters ($\varepsilon =0.01$,$\lambda=0.2$) to
show the good fitting. The results are within 1$\sigma$
expectation.}
\end{figure}


\begin{thebibliography}{99}


\bibitem{pr} S. Perlmutter et al., Astrophys. J. {\bf 517} (1999) 565-586,
astro-ph/9812133; A. G. Riess et al., Astron. J. {\bf 116} (1998)
1009-1038, astro-ph/9805201; Astrophys.J. 560 (2001) 49-71,
astro-ph/0104455.

\bibitem{turner} M. S. Turner, astro-ph/0202008 (To appear in the
Proceedings of 2001: A Spacetime Odyssey (U. Michigan, May 2001,
World Scientific)).

\bibitem{faraoni} V. Faraoni, Int. J. Mod. Phys. D{\bf 11} (2002)
471-482, astro-ph/0110067; E. Gunzig, A. Saa, L. Brenig, V.
Faraoni, T.M. Rocha Filho and A. Figueiredo, Phys. Rev. D{\bf 63}
(2001) 067301, gr-qc/0012085.

\bibitem{uzan} A. Riazuelo, J.-P. Uzan, Phys. Rev. D{\bf 66}
(2002) 023525, astro-ph/0107386; J.-P. Uzan, Phys. Rev. D{\bf 59}
(1999) 123510, gr-qc/9903004; F. Perrotta, C. Baccigalupi,
astro-ph/0205245, Phys. Rev. D{\bf 65} (2002) 123505,
astro-ph/0201335; F. Perrotta, C. Baccigalupi, S. Matarrese,
 Phys. Rev. D{\bf 61} (2000) 023507, astro-ph/9906066; T. Chiba,
 Phys. Rev. D{\bf 64} (2001) 103503, astro-ph/0106550.

\bibitem{dftorres} D. F. Torres, Phys. Rev. D{\bf 66}
(2002) 043522, astro-ph/0204504.

\bibitem{peebles} P. J. E. Peebles and Bharat Ratra, astro-ph/0207347.

\bibitem{ratrap} B. Ratra and P. J. E. Peebles, Phys. Rev.
D{\bf 37} (1988) 3406.

\bibitem{chimento} L. P. Chimento and A. S. Jakubi, Int. J.
Mod. Phys. D{\bf 5} (1996) 71-84, gr-qc/9506015.

\bibitem{starobinsky} A. A. Starobinsky, Grav. Cosmol. {\bf 4}
(1998) 88-99, astro-ph/9811360.

\bibitem{rs}  C. Rubano and P. Scudellaro, Gen. Rel. Grav. {\bf 34}
(2002) 307-328, astro-ph/0103335.

\bibitem{tmatos} L. A. Urena-Lopez, T. Matos, Phys. Rev. D{\bf 62}
(2000) 081302, astro-ph/0003364.

\bibitem{bcn} T. Barreiro, E. J. Copeland and N. J. Nunes,
Phys. Rev. D{\bf 61} (2000) 127301, astro-ph/9910214; E. J.
Copeland, N. J. Nunes, F. Rosati, Phys. Rev. D{\bf 62} (2000)
123503, hep-ph/0005222.

\bibitem{ellis} G. F. R. Ellis and M. Madsen, Class. Quant.
Grav. {\bf 8} (1991) 667.

\bibitem{uggla} C. Uggla, R. T. Jantzen and K. Rosquist,
Gen. Rel. Grav. {\bf 25} (1993) 409.

\bibitem{sen} A. A. Sen, S. Sethi, Phys. Lett. B{\bf 532}
(2002) 159-165, gr-qc/0111082.

\bibitem{saini} T. D. Saini, S. Raychaudhury, V. Sahni and
A. A. Starobinsky, Phys. Rev. Lett. {\bf 85} (2000) 1162-1165,
astro-ph/9910231.

\bibitem{ygong} Y. Gong, Class. Quant. Grav. {\bf 19}
(2002) 4537-4542, gr-qc/0203007.

\bibitem{jdbarrow}  C. Rubano and J. D. Barrow, Phys. Rev.
D{\bf 64} (2001) 127301, gr-qc/0105037.

\bibitem{holden} D. J. Holden and D. Wands, Phys. Rev., D{\bf 61}, 043506 (2000),
gr-qc/9908026.

\bibitem{chimento1} L. P. Chimento, Class. Quantum Grav.
{\bf 15} (1998) 965-974; J. M. Aguirregabiria and L. P. Chimento,
Class. Quantum Grav. {\bf 13} (1996) 3197-3209; L. P. Chimento ,
A. E. Cossarini and N. A. Zuccala; Class. Quantum Grav. {\bf 15}
(1998) 57-74; L. P. Chimento, N. A. Zuccala and V. Mendez; Class.
Quantum Grav. {\bf 16} (1999) 3749-3763.

\bibitem{acq} O. Arias, R. Cardenas, I. Quiros, Nucl. Phys.
B{\bf 643} (2002) 187-200, hep-th/0202130.

\bibitem{clw} E. J. Copeland, A. R. Liddle and D. Wands,
gr-qc/9711068.

\bibitem{will} C. M. Will, {\it Theory and Experiment in
Gravitational Physics} (Cambridge University Press, 1993);
gr-qc/0103036.

\bibitem{triess} A. Riess, astro-ph/0104455; M. S. Turner
and A. Ries, astro-ph/0106051 (Astrophys. J., in press).

\bibitem{ptw} S. Perlmutter, M. S. Turner and M. White,
Phys. Rev. Lett. {\bf 83} (1999) 670-673, astro-ph/9901052; M. S.
Turner and M. White, Phys. Rev. D{\bf 56} (1997) R4439.

\bibitem{cc} P. S. Corasaniti and E. J. Copeland, Phys. Rev. D{\bf 65} (2002)
043004, astro-ph/0107378.

\bibitem{bhm} R. Bean, S. H. Hansen, and A. Melchiorri, Nucl. Phys. Proc. Suppl.
{\bf 110} (2002) 167 , astro-ph/0201127.

\bibitem{mst} M. S. Turner, Nucl. Phys. Proc. Suppl. {\bf 72} (1999)
69, astro-ph/9811366.

\bibitem{rubano}  M. Pavlov, C. Rubano, M. Sazhin and P.
Scudellaro, Astrophys.J. {\bf 566} (2002) 619, astro-ph/0106068.

\bibitem{perlmutter} S. Perlmutter, G. Aldering, G. Goldhaber et
al., ApJ {\bf 517} (1999) 565.


\end{thebibliography}
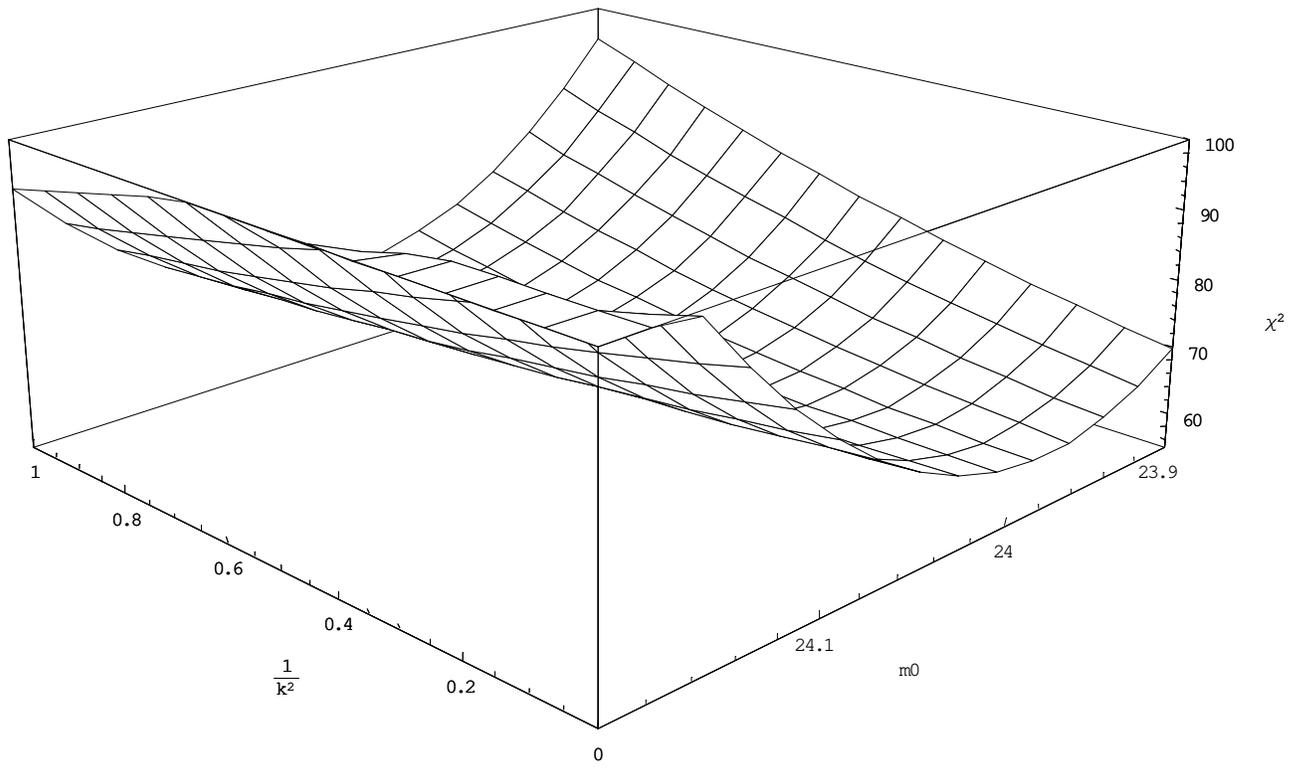
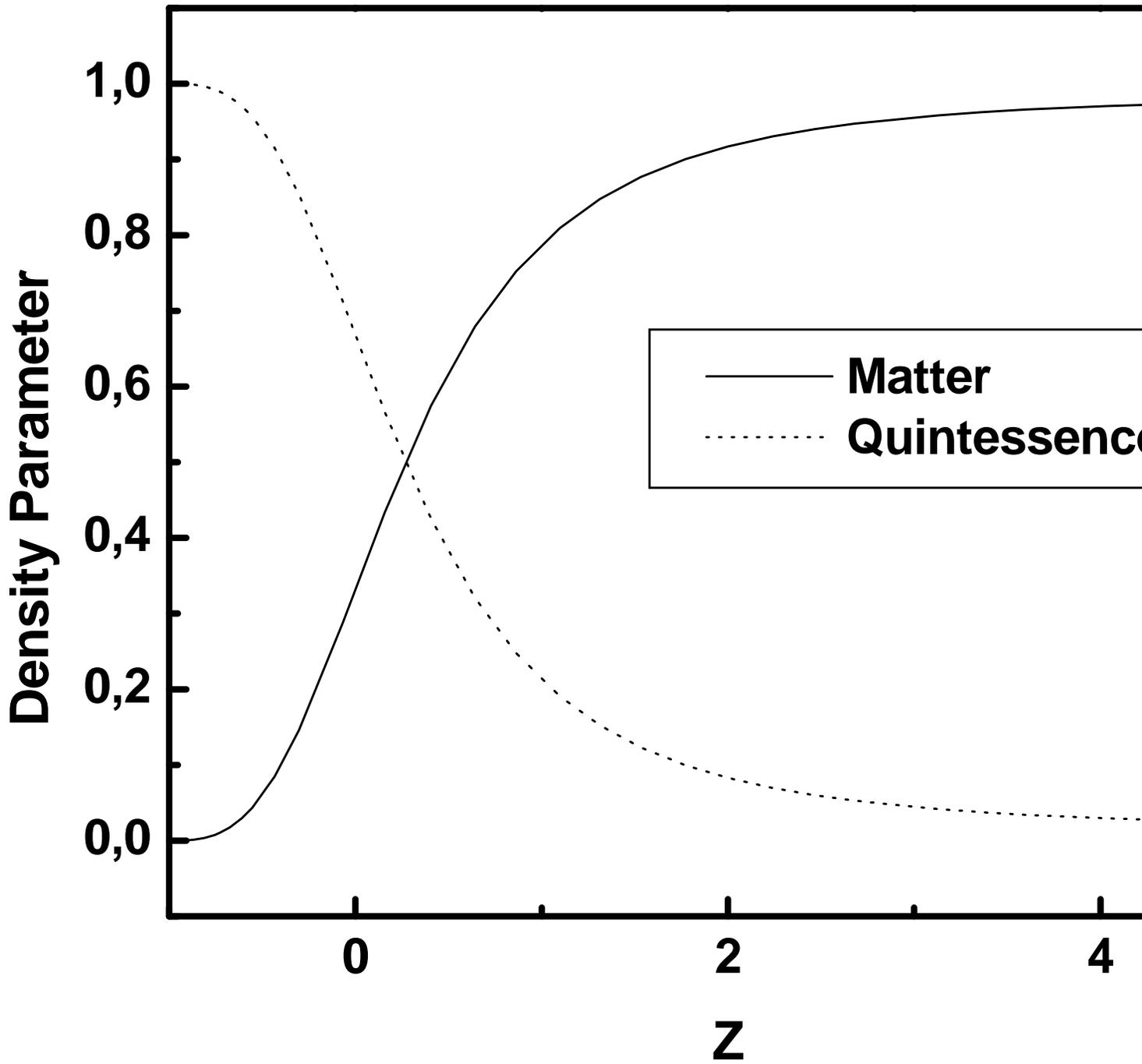
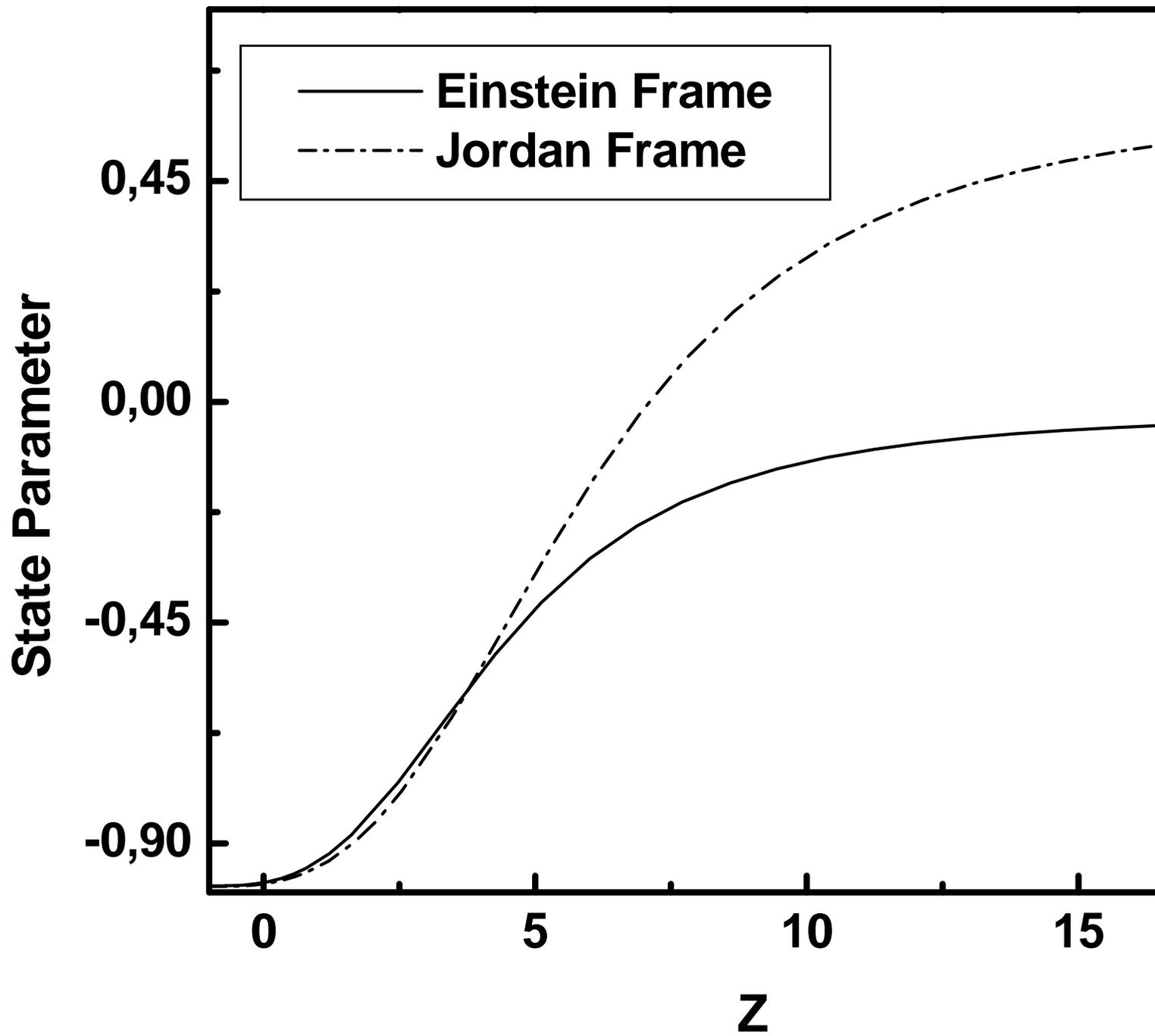
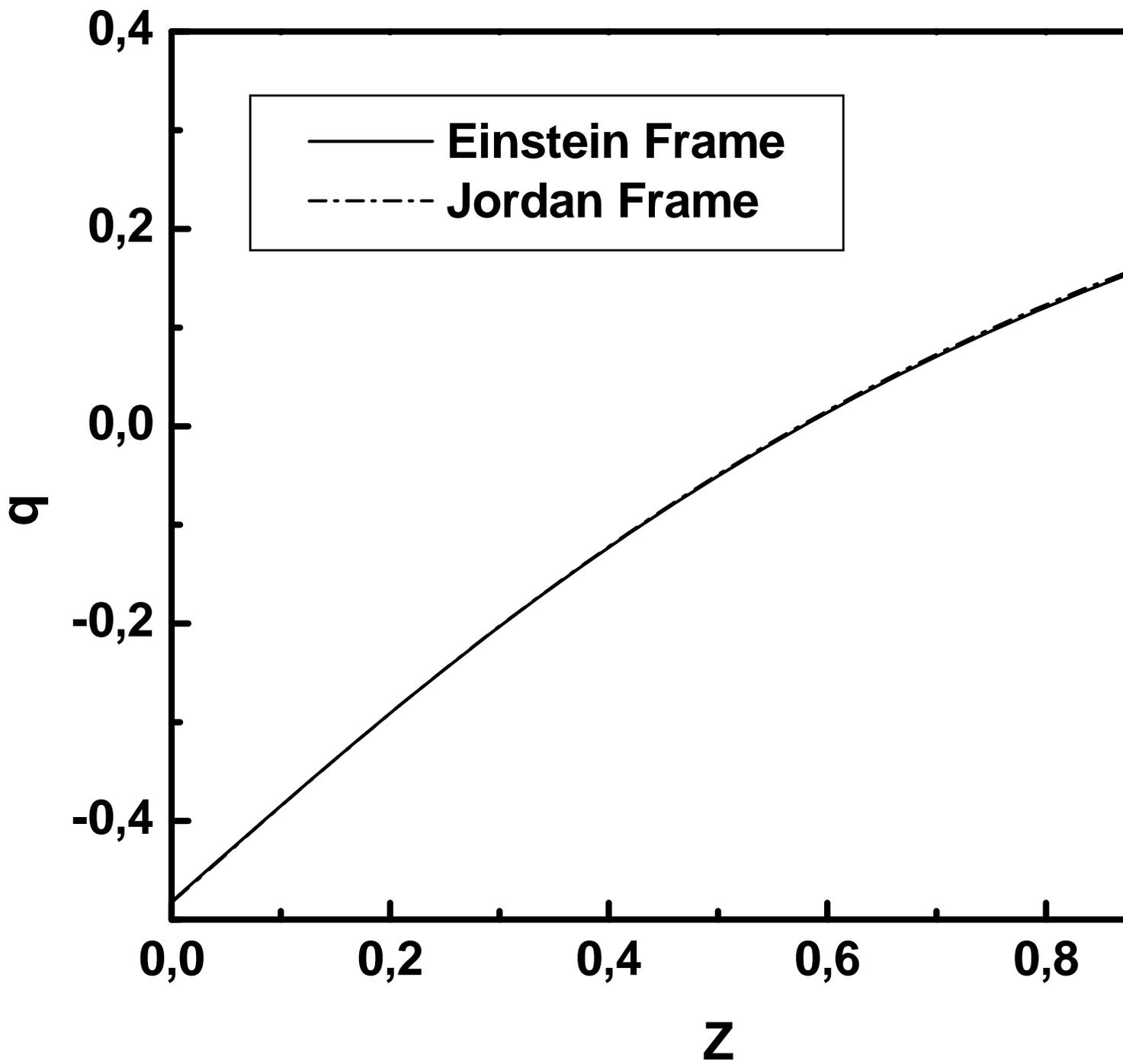
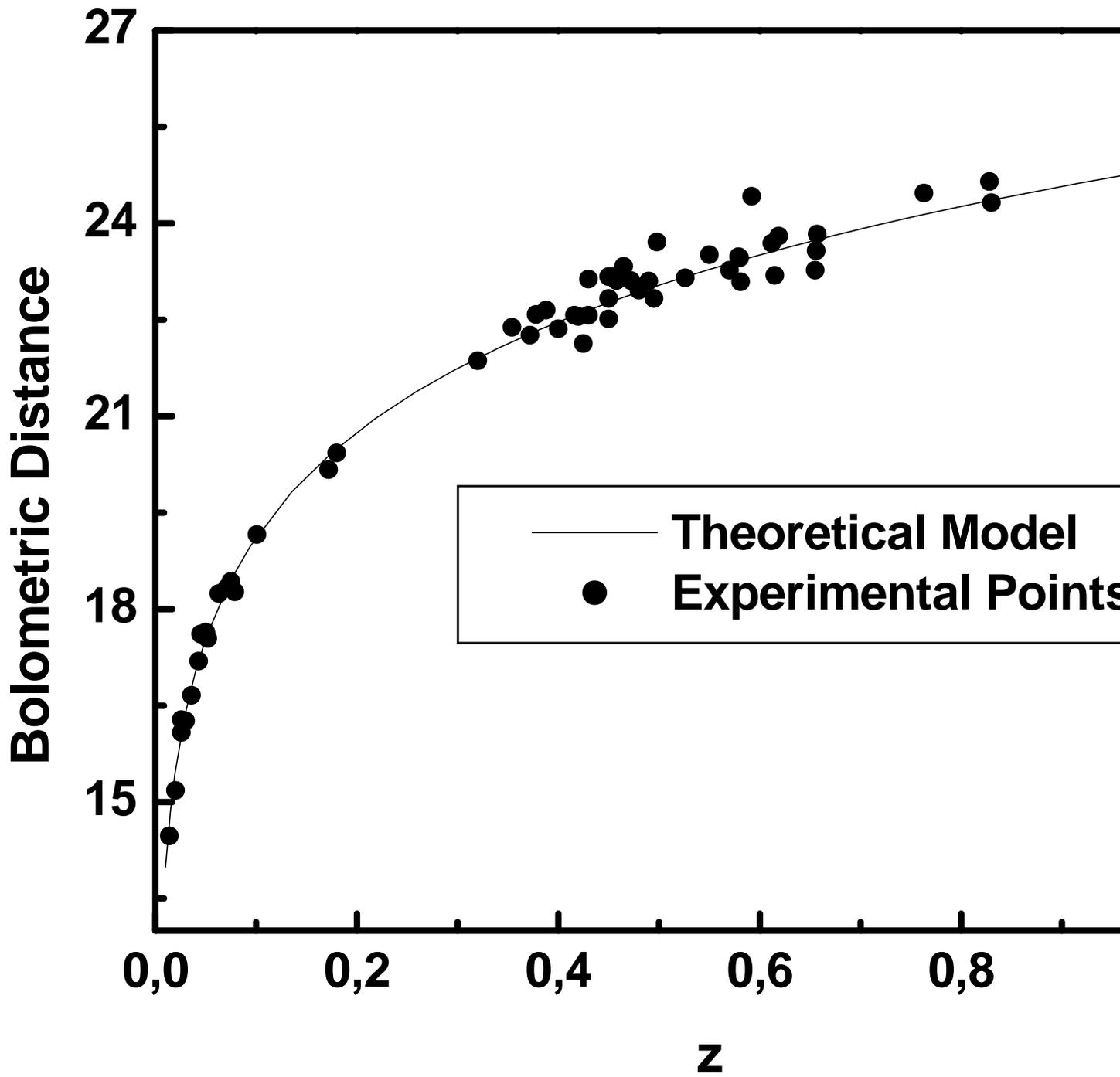
\end{document}